\documentclass[twocolumn]{revtex4}

\usepackage{psfrag}
\usepackage{amssymb}
\usepackage{indentfirst}
\usepackage[dvips]{graphicx}
\usepackage{psfig}

\begin{document}

\title{Radiative heat transfer between metallic nanoparticles}
\author{Pierre-Olivier Chapuis$^{1,2}$\footnote{Electronic mail: olivier.chapuis@centraliens.net}}
\author{Marine Laroche$^{1}$}
\author{Sebastian Volz$^{1}$}
\author{Jean-Jacques Greffet$^{1}$}
\affiliation{$^{1}$Laboratoire d'Energ\'etique Mol\'eculaire et Macroscopique, Combustion\\CNRS UPR 288, Ecole Centrale Paris\\Grande Voie des Vignes, F-92295 Ch\^atenay-Malabry cedex, France\\
$^{2}$Institut des NanoSciences de Paris\\
Universit\'e Pierre et Marie Curie-Paris 6, CNRS UMR 7588, Universit\'e Denis Diderot-Paris 7\\
Campus Boucicaut, 140 rue de Lourmel, F-75015 Paris, France
}
\linespread{1}
\begin{abstract}
In this letter, we study the radiative heat transfer between two nanoparticles in the near field and in the far field. We find that the heat transfer is dominated by the electric dipole-dipole interaction for dielectric particles and by the magnetic dipole-dipole interaction for metallic nanoparticles. We introduce polarizabilities formulas valid for arbitrary values of the skin depth. While the heat transfer mechanism is different for metallic and dielectric nanoparticles, we show that the distance dependence is the same. However, the dependence of the heat flux on the particle radius is different. \end{abstract}
\maketitle

In a vacuum, the heat flux between two bodies is only due to radiative heat transfer. It was predicted four decades ago~\cite{Polder} that the radiative heat flux exchanged by two bodies increases dramatically when the distance between them decreases down to distances on the order of the peak wavelength of Planck's spectrum. The heat transfer between two nanoparticles has been theoretically predicted by modeling the nanoparticles using electric dipoles~\cite{Volokitin2001,Domingues,Dorofeyev}. Experiments between macroscopic media~\cite{Tien,Hargreaves} or involving a nano-object~\cite{Kittel} have been reported.
Different behaviors for polar materials and metallic ones were found for the interaction of a surface and a particle~\cite{Chapuis}.
In this letter, we show that the heat power exchange between metallic nanoparticles is dominated by the magnetic dipole contribution. We derive formulas valid for arbitrary skin depths values. We also show that the heat transfer has the same dependence with the interparticle distance $d$ for polar or metallic particles. It decays as $1/d^6$ in the near field and as $1/d^2$ in the far field. We find that the radius dependence differs for metallic and dielectric nanoparticles.

\begin{figure}[h]
\begin{center}
\includegraphics[width=4.5cm]{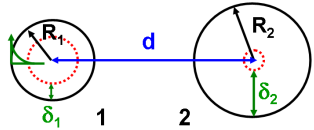}
\caption{Two nanoparticles with radii $R_{1}$ and $R_{2}$ at an interparticular distance $d$. The skin depth $\delta_{1}$ and $\delta_{2}$ are fundamental quantities for metallic particles.}
\label{Fig:Intro}
\end{center}
\end{figure}
The derivation of the radiative heat flux $\Phi=\Delta P$ exchanged by two nanoparticles is done in the framework of fluctuational electrodynamics~\cite{SurfaceScienceReports,ReviewVolokitin}. It is based on the absorption by a particle (P1) of the electromagnetic fields generated by the random currents in the other particle (P2) and vice versa.
The power absorbed by P1 reads $
P=\int d^3\vec{r} <\vec{j}(\vec{r},t)\vec{E}_{int}(\vec{r},t)>
$ where $\vec{j}$ is the electric current density in P1 and $\vec{E}_{int}$ the electric field at point $\vec{r}$ in this particle. We consider a particle with radius R such that $R \ll \lambda$, where the vacuum wavelength $\lambda$ is in the infrared. $R$ could be on the same order of magnitude than the skin depth, which is defined by $\delta(\omega)=1/[{\mbox Im}(\sqrt{\varepsilon}) k]$, where $k=2\pi / \lambda=\omega/c$ and $c$ is the light velocity. $\varepsilon(\omega)$ is the relative dielectric constant of the medium. In the IR frequency range, the media are non-magnetic.

The power absorbed by a particle illuminated by a plane wave is the product of the incident flux by the particle absorption cross section. The latter reads~\cite{Bohren} $
C_{abs}=C_{ext}-C_{sca}$,
where we have introduced the extinction cross section $C_{ext}$ and the scattering cross section $C_{sca}$. A first order expansion in $x=kR$ yields
\begin{eqnarray}
C_{abs}=\frac{6\pi}{k^2}~{\mbox Re}(a_1+b_1).
\end{eqnarray}
Note that this expansion is done for arbitrary values of $y=\sqrt{\varepsilon(\omega)}~x$. $a_1$ and $b_1$ are the first Mie coefficients. They are related respectively to the electric and magnetic dipolar moment~\cite{Bohren,Mulholland}. The power absorbed by the particle can be cast in the form
\begin{eqnarray}
P=<\frac{\partial \vec{p}}{\partial t} \cdot \vec{E}_{inc}-\vec{m} \cdot \frac{\partial \vec{B}_{inc}}{\partial t}>
\label{eq:generale}
\end{eqnarray}
where $\vec{E}_{inc}$($\vec{H}_{inc}$) are the incident electric (magnetic) fields and where $\vec{p}$ and $\vec{m}$ are the electric and magnetic dipolar moment of the particle. The dipolar approximation is valid provided that $\lambda \gg R$ and $d \gg R$, where $d$ is the distance between the center of the nanoparticles. In practice, a distance on the order of a few radii (8) appears to be sufficient~\cite{Domingues}. For smaller distances higher multipoles must be included~\cite{Dorofeyev,Arvind,Rubi}.

For metallic nanoparticles, it has been shown~\cite{Chapuis,Dedkov} that the magnetic dipole moment gives a significant contribution to absorption in the near field. These works focus on very small nanoparticles ($R\ll \delta $)~\cite{Chapuis} or 
larger ones ($R\gg \delta $)~\cite{Dedkov}. As particle sizes are generally not smaller or larger than the skin depth in the full contributing spectrum, an extended formula valid for arbitrary-large skin depths is needed. Provided that $R \ll \lambda$, we derive from Mie's solution the following magnetic polarizability
\begin{eqnarray}
\label{Eq3}
\alpha_H(\omega) =&-&2 \pi 
R^3\mbox{\huge{[}}
\left(1-\frac{x^2}{10} \right) \\
\nonumber
&+& \left( -\frac{3}{y^2}
+\frac{3}{y} \mbox{cotan} y \right) \left( 1-\frac{x^2}{6} 
\right)   \mbox{\huge{]}}
\end{eqnarray}
In the limit $R\ll \delta$, Eq.(3) yields the magnetic dipolar moment valid for very small particles
\begin{equation}
\alpha_H(\omega) =\frac{2\pi}{15}R^3(kR)^2[\varepsilon(\omega)-1].
\label{bleu}
\end{equation}

The electric polarizability takes the simple Clausius-Mossotti form $
\alpha_E(\omega)=4\pi R^3 \frac{\varepsilon(\omega)-1}{\varepsilon(\omega)+2}
$
if the skin depth is much larger than the radius. For dielectric nanometer scale particles, this is often a very good approximation. Yet, the skin depth effect may play a role. 
It is then useful to work with a more general form derived from Mie solution:
\begin{eqnarray}
\nonumber
&&\alpha_E(\omega) =2 \pi 
R^3
\mbox{\huge{[}} 2\left( \mbox{sin} y -y \mbox{cos} y \right) \\
\nonumber
&&-x^2 \left( \frac{-\mbox{sin} y}{y^2} +  \frac{\mbox{cos} y}{y} + \mbox{sin} y  \right) \mbox{\huge{]}}/
\mbox{\huge{[}} \left(  \mbox{sin} y -y \mbox{cos} y \right)\\
&&+ x^2\left( \frac{-\mbox{sin} y}{y^2} +  \frac{\mbox{cos} y}{y} + \mbox{sin} y  \right) \mbox{\huge{]}}.
\label{eq:bab}
\end{eqnarray}
Note that the famous Clausius-Mossotti formula is recovered in the limit $R \ll \delta$.
We have represented in Fig. 2 the imaginary part of the electric and magnetic polarizabilities for different values of the radius of a gold nanoparticle. It is seen that the simplest forms are valid for any frequency when the radius is smaller than 20 nm.

\begin{figure}[h]
\begin{center}
\includegraphics[width=7.5cm]{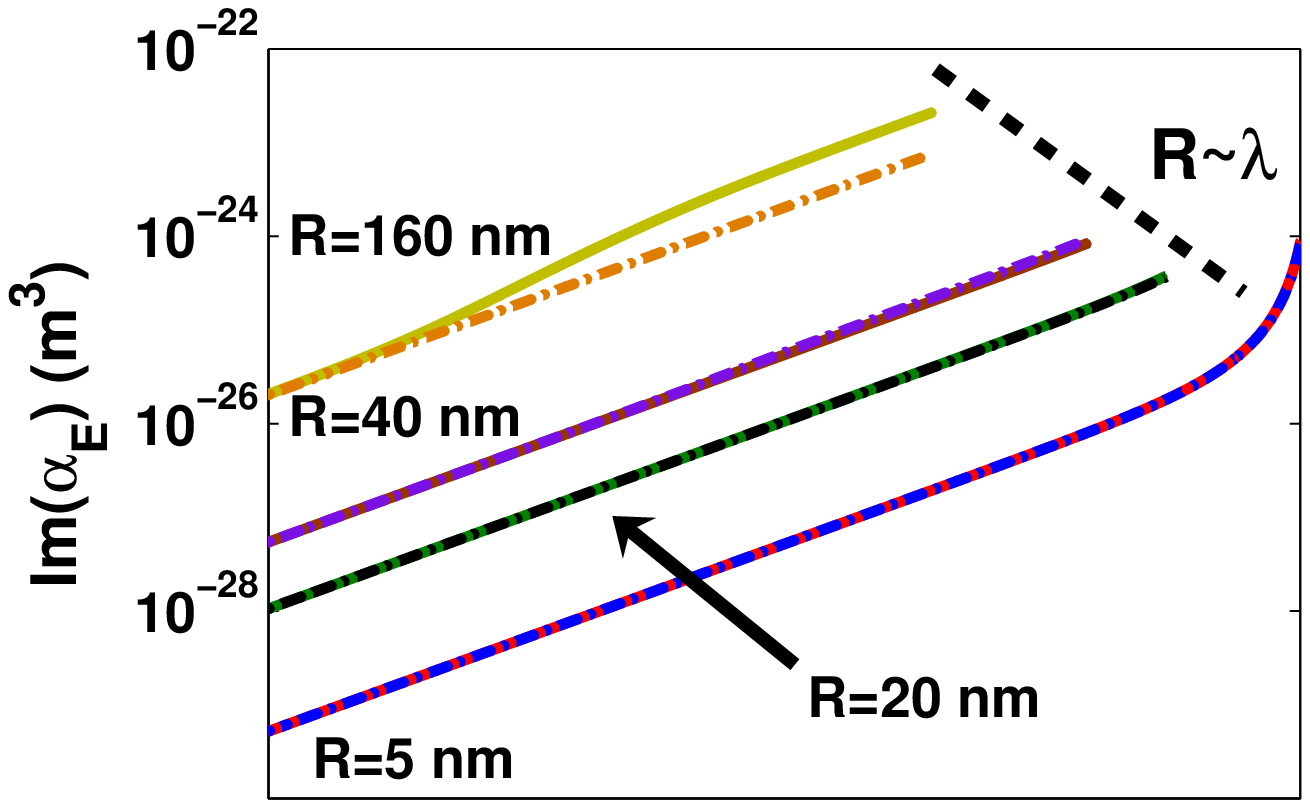}
\includegraphics[width=7.5cm]{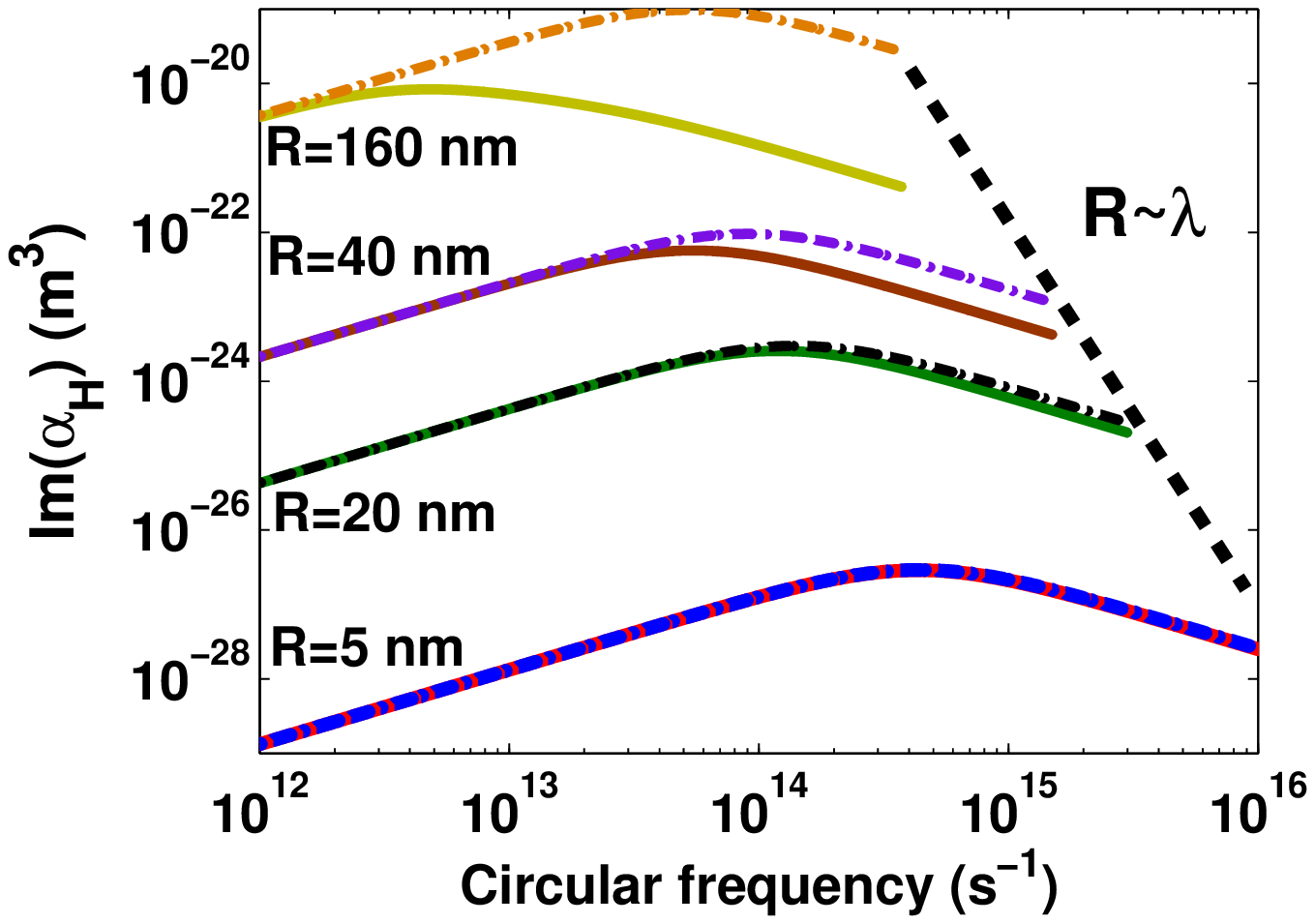}
\caption{Imaginary parts of the electric and magnetic polarizabilities for different gold particle radii. Plain lines are obtained by considering Eq. (\ref{Eq3}) or (\ref{eq:bab}) and dotted lines are obtained by applying the approximations given by Eq. (\ref{bleu}) or the Clausius-Mossotti formula.
}
\label{fig:polar}
\end{center}
\end{figure}

Eq. (\ref{eq:generale}) leads to
\begin{eqnarray}
P=\int_{\omega=0}^{+\infty}   
d\omega
~\omega \mbox{ \huge{[}}
Im \left[  \alpha_E (\omega) \right] \varepsilon_0 <\left|   \vec{E}^{inc} \right|^2> \\
\nonumber
+Im \left[ \alpha_H(\omega) \right]      \mu_0    <  \left| \vec{H}^{inc} \right|^2> \mbox{ \huge{]}}
\label{Eq:totale}
\end{eqnarray}
where $\varepsilon_0$ and $\mu_0$ are the free space permittivity and permeability.
The incident fields are generated by the fluctuating dipoles. The fluctuation-dissipation theorem~\cite{SurfaceScienceReports,Landau,Agarwal} yields:
\begin{eqnarray}
&<p_{m}(\vec{r},\omega)~p_{\ell}^{*}(\vec{r},\omega ')>\\
\nonumber
&=\frac{4\pi\epsilon_0}{\omega} Im[\alpha_E] \Theta (\omega,T) \delta(\omega-\omega')\delta_{m\ell}
\end{eqnarray}
where $\Theta (\omega,T)=\hbar \omega/[exp(\hbar \omega/k_B T)-1]$ is the mean energy of a mode ($\hbar$ and $k_B$ are the Planck and Boltzmann constants). In a non-magnetic medium ($\mu=1$), the magnetic dipoles satisfy:
\begin{eqnarray}
&<m_{m}(\vec{r},\omega)~m_{\ell}^{*}(\vec{r},\omega ')>\\
\nonumber
&=\frac{4\pi}{\omega \mu_0} Im[\alpha_H] \Theta (\omega,T) \delta(\omega-\omega')\delta_{m\ell}.
\end{eqnarray}
Using a Green's function technique~\cite{SurfaceScienceReports,Aussois} relating the source dipoles to the emitted fields, we obtain
\begin{eqnarray}
&&\Phi=\sum_{a=H~or~E} \int_0^{+\infty} \frac{1}{4 \pi^3} \\
\nonumber
&&[\Theta(\omega,T_1)-\Theta(\omega,T_2)]~\mbox{Im}\left(\alpha_a^1(\omega)\right)\mbox{Im}\left(\alpha_a^2(\omega)\right)\\
\nonumber
&&k^6\left[ \frac{3}{(kd)^6}+\frac{1}{(kd)^4}+\frac{1}{(kd)^2}\right]d\omega,
\label{eq:waouhh}
\end{eqnarray}
where $\Theta(\omega,T)=\hbar \omega /[exp(\hbar \omega/k_B T)-1]$ is the mean energy of an oscillator. We checked that multiple scattering is negligible for sufficiently small particles. It is not taken into account in Eq. (9).

\begin{figure}[h]
\begin{center}
\includegraphics[width=7.5cm]{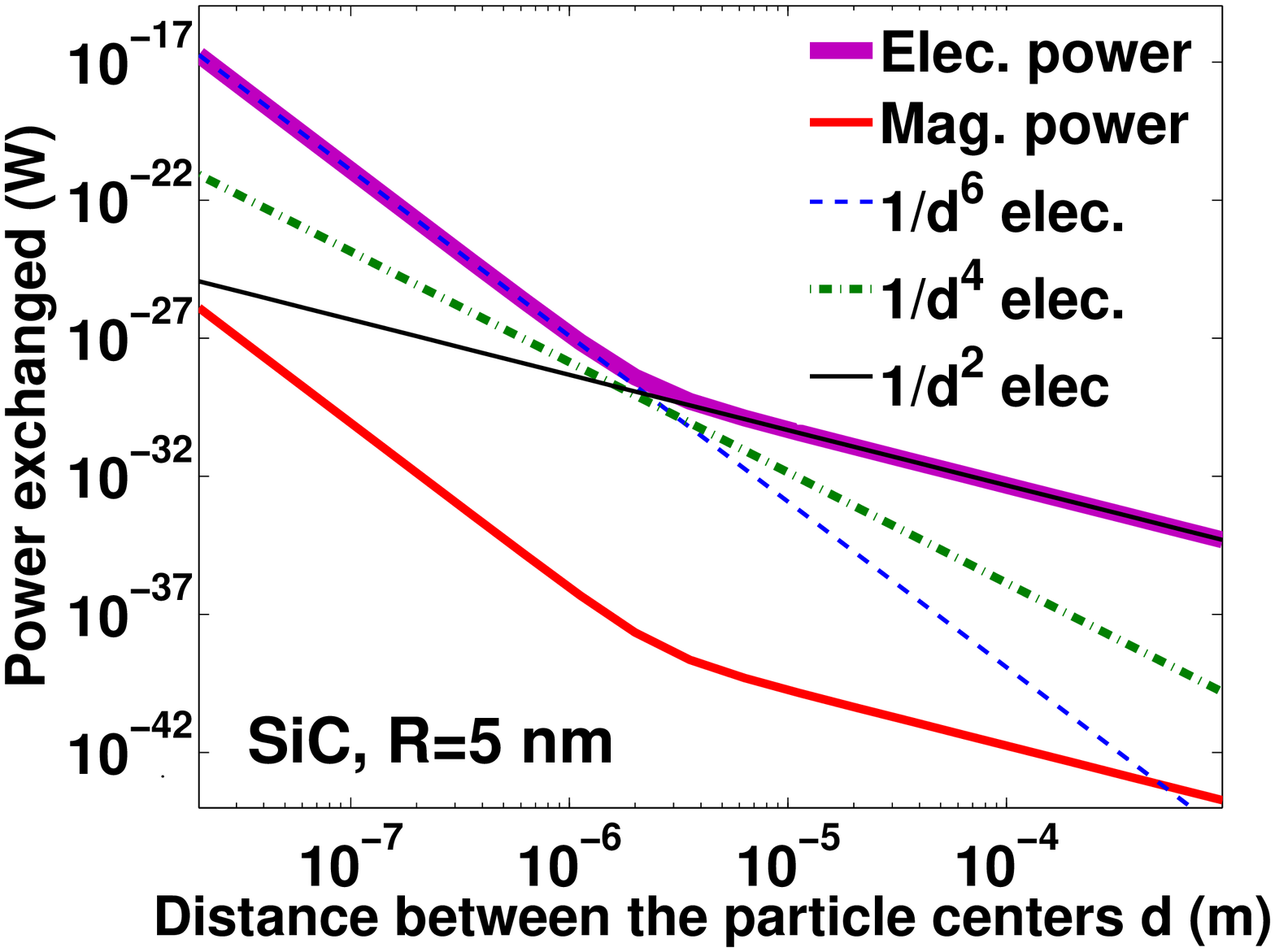}
\includegraphics[width=7.5cm]{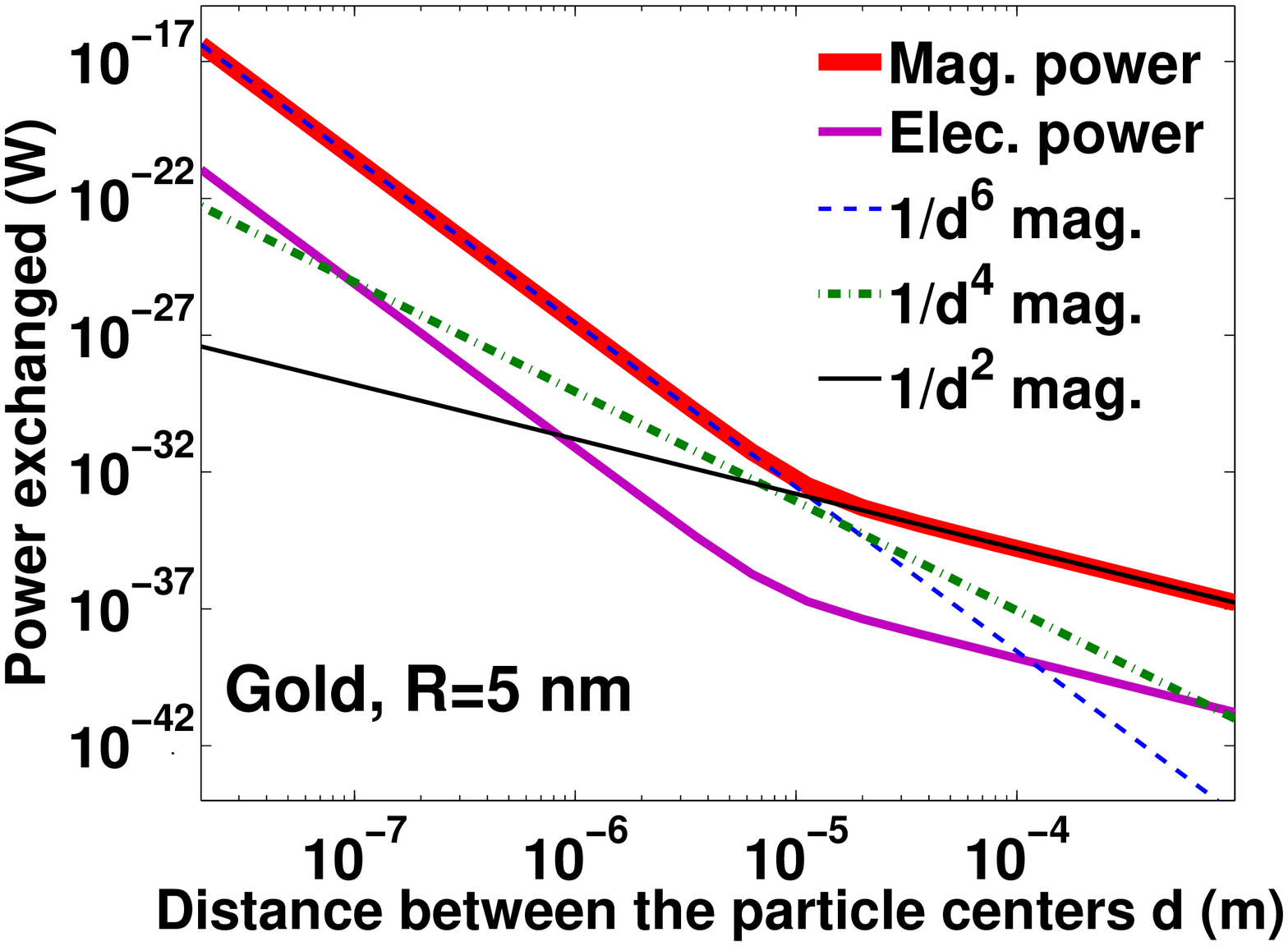}
\caption{Radiative power exchange between two SiC and gold nanoparticles of $5$~nm radii, one being at $300$~K and the other at $400$~K. Dielectric-particles relative permittivities are assimilated to a Lorenz model $\varepsilon=\varepsilon_{\infty}\left[1+(\omega_L^2-\omega_T^2)/(\omega2-\omega_T^2-i\Gamma \omega)\right]$\cite{Mulet,Palik} 
and metallic-particles ones are assimilated to a modified Drude model 
$\varepsilon=1-\omega_{P}^{2}/\{\omega[\omega+i (\nu_0+A~v_F/R)]\}$.
$v_F$, the Fermi velocity, and A, a constant on the order of unity, account for confinement effects (see Ref. \cite{Chapuis} for parameters)
.}
\label{fig:bob}
\end{center}
\end{figure}
The final result has the same structure for the electric and magnetic dipolar moments. 
It shows that the heat flux dependence on distance is the same for polar and metallic materials. This is in contrast with other studies with larger bodies, e.g. parallel surfaces~\cite{ChapuisNonLocal} or nanoparticle/surface~\cite{Chapuis,Dedkov}.

We plot in Fig. \ref{fig:bob} the heat transfer between two dielectric (SiC) and two metallic nanoparticles. It is seen that the electric dipole yields the dominant contribution for dielectric particles whereas the magnetic contribution dominates the heat transfer for gold nanoparticles, for the whole range of distances. Heat transfer between metallic nanoparticles is thus mainly due to Joule dissipation of eddy currents. A quasistatic approximation neglecting magnetic fields is valid for dielectric particles but not for metallic particles. It is also worth noting that the $1/d^4$ term is almost negligible. The near-field term decays as $1/d^6$ and the far-field one as $1/d^2$. To first approximation, one can retain the sum of these two contributions.

We now discuss the dependence of the heat transfer on the particle radius. The flux varies as the square of the polarizability so that it varies as $R^6$ for dielectric particles that have the same radii. The dependence is different for the magnetic contribution that dominates the flux for metallic particles. At first sight, the polarizability varies as $R^5$ from Eq. (4). 
However, for the very small particles, the confinement effect introduces a further dependence of the dielecric constant on $R$ (see Ref. \cite{Tomchuk}). We have $\mbox{Im}(\varepsilon) \approx R~\omega_p^2/(\omega v_F)$. It follows that the flux varies as $R^{12}$. This dependence becomes weaker when the radius increases.

In summary, we have given here a simple and general formula for the near-field radiative heat transfer between two nanoparticles, valid for $\lambda \gg R $, $d \gg R $ and any value of the skin depth. We have shown that heat transfer is dominated by magnetic fields for metallic nanoparticles and by electric fields for dielectric nanoparticles. Research on heat transfer properties of composite media including nanoparticles should benefit from this work.


The authors acknowledge the support of \textit{Agence Nationale de la Recherche} through the \textit{Ethna}, \textit{ThermaEscape} and \textit{Monaco} projects. P.O.C. thanks B. Palpant for discussions.

{
\small
\linespread{1}

}

 \end{document}